\documentclass[manuscript]{emulateapj}
\usepackage{natbib}
\usepackage{epsfig}
\usepackage{lscape}
\usepackage{rotating}
\usepackage{lscape,graphicx}

\usepackage{color}

\newcommand{\cmjj}{\mbox{${\rm cm^{-2}}$}}

\newcommand{\lya}{\mbox{${\rm Ly}\alpha$}}

\newcommand{\apg}{\gtrsim}
\newcommand{\apll}{\lesssim}
\newcommand{\etal}{\ensuremath{\mbox{et~al.}}}
\newcommand{\hmsol}{\mbox{$h^{-1}\,{\rm M}_\odot$}}
\newcommand{\hkpc}{\mbox{$h^{-1}~{\rm kpc}~$}}
\newcommand{\hmpc}{\mbox{$h^{-1}~{\rm Mpc}~$}}
\providecommand{\kms}{\,\ensuremath{\rm{km\,s}^{-1}}}

\slugcomment{Accepted for publication in the Astrophysical Journal}

\shorttitle{Incidence of cool gas in $10^{13}$ halos}
\shortauthors{Gauthier et al.}

\begin{document}

\title{The incidence of cool gas in $\sim$ $10^{13}~{\rm M}_{\odot}$ halos$^1$}

\author{Jean-Ren\'e Gauthier\altaffilmark{2,3}, Hsiao-Wen Chen\altaffilmark{2} and Jeremy L. Tinker\altaffilmark{4}}

\altaffiltext{1}{ This paper includes data gathered with the 2.5 meter du Pont telescope located at Las Campanas 
Observatory, Chile and with the Apache Point Observatory 3.5-meter telescope, which is owned and operated 
by the Astrophysical Research Consortium.  }
\altaffiltext{2}{Department of Astronomy \& Astrophysics and Kavli Institute for Cosmological Physics, University of Chicago, IL; gauthier@oddjob.uchicago.edu}
\altaffiltext{3}{The Observatories of the Carnegie Institution for Science, 813 Santa Barbara Street, Pasadena, CA, 91101}
\altaffiltext{4}{Berkeley Center for Cosmological Physics, University of California, Berkeley, CA}

\begin{abstract}

We present the first results of an ongoing spectroscopic follow-up of
close luminous red galaxy (LRGs) and Mg\,II $\lambda\lambda$ 2796,2803
absorber pairs for an initial sample of 15 photometrically selected
LRGs at physical projected separations $\rho \apll 350$ \hkpc\ from a
QSO sightline.  Our moderate-resolution spectra confirm a physical
association between the cool gas ($T \sim 10^4$ K) revealed by the
presence of Mg\,II absorption features and the LRG halo in five cases.
In addition, we report an empirical estimate of the maximum covering
fraction ($\kappa_{\rm max}$) of cool gas in massive, $\apg 10^{13}$
\hmsol\ dark matter halos hosting LRGs at $z \sim 0.5$. This study is
performed using a sample of foreground LRGs that are located at $\rho
< 400$ \hkpc\ from a QSO sightline.  The LRGs are selected to have a
robust photometric redshift $\sigma_z/(1+z_{\rm ph})\simeq 0.03$.  We
determine $\kappa_{\rm max}$ based on the incidence of Mg\,II
absorption systems that occur within $z_{\rm ph} \pm 3\,\sigma_{z}$ in
the spectra of the background QSOs.  Despite the large uncertainties in
$z_{\rm ph}$, this experiment provides a conservative upper limit to
the covering fraction of cool gas in the halos of LRGs. We find that
$\kappa_{\rm max} \approx 0.07$ at $W_r(2796)\ge 1.0$ \AA\ and
$\kappa_{\rm{max}} \approx 0.18$ at $W_r(2796)\ge 0.5$ \AA, averaged
over 400 \hkpc\ radius.
Our study shows that while cool gas is present in $\apg$ $10^{13}$ \hmsol\
halos, the mean covering fraction of strong absorbers is no more than
7\%.

\end{abstract}


\section{Introduction}

A detailed description of how galaxies acquire their gas is essential
to establish a comprehensive theory of galaxy evolution. Analytical
calculations and hydrodynamical simulations have shown that the gas in
$M\apg 10^{12}$ M$_{\odot}$ halos\footnote{ We define a halo as a
region of overdensity
200 with respect to the mean mass density of the universe.},
is heated to high temperature by virial shocks (e.g.,\
\citealt{birnboim2003a,keres2005a,dekel2006a,keres2009b}). Within the
hot gas halo, thermal instabilities can induce the formation of
pressure-supported clouds of cool gas (e.g.,\
\citealt{mo1996a,maller2004a}). Such clouds have been resolved in
high-resolution hydrodynamic simulations of Milky-Way type dark matter
halos (e.g.,\ Kauffmann et al.\ 2008; Kere\v{s} \& Hernquist
2009). These clouds bring some baryonic material into the galaxy, but
the overall cooling rate is reduced because the hot phase has lower
density. Whether or not these clouds can form and survive in more
massive halos is an open question.

Our empirical knowledge of the cool gas content of dark matter halos
has been shaped primarily by H\,I 21-cm observations of local galaxies
(e.g.,\ \citealt{thilker2004a, doyle2005a, verdes-montenegro2001a,
verdes-montenegro2007a}). At $z\apg0.1$, we rely on intervening
systems detected along the sightlines to background QSOs to describe
the gas content of dark matter halos.  (e.g.,\ \citealt{lanzetta1990a,
steidel1994a, chen2001b, chen2008a}).  A number of studies have used
the Mg\,II $\lambda\lambda$ 2796,2803 absorption doublets to trace
cool, $T \sim 10^4$ K, gas and its physical association with
foreground galaxies located near QSO sightlines (e.g.,\
\citealt{tripp2005a,kacprzak2008a,chen2008a}).  Insights into the cool
gas content of dark matter halos can also be obtained by measuring the
large-scale clustering amplitude of Mg\,II systems
\citep{bouche2006a,tinker2008a,tinker2010a}.
 
In \citet{gauthier2009a}, we calculated the clustering amplitude of
strong Mg\,II absorbers (with rest-frame absorption equivalent width
$W_r(2796)\ge 1$ \AA) using luminous red galaxies (LRGs) at $z\sim
0.5$. LRGs are old and passive galaxies \citep{eisenstein2001a}
residing in $M \apg 10^{13}$ \hmsol\ halos. (e.g.,\
\citealt{zheng2008a,blake2008a,padmanabhan2008a,gauthier2009a}).  They
are identified using photometric redshift techniques that offer a
typical redshift accuracy of $\sigma_z \approx 0.045$ at $z_{\rm ph}
\sim 0.5$ \citep{collister2007a}.  

An interesting result from
\citet{gauthier2009a} is that the LRG-Mg\,II cross-correlation signal
is comparable to the LRG auto-correlation on small scales ($\apll 300$
co-moving \hkpc) that are well within the virial radii of the
halos. The results suggest the presence of cool
gas inside the dark matter halos of LRGs. 

Here we report the first results of an ongoing spectroscopic follow-up
of the close LRG--Mg\,II pairs found in \citet{gauthier2009a}.  In
five of the 15 cases, the precise spectroscopic redshifts establish
a physical connection between these LRG--Mg\,II pairs.  The
spectroscopic study of LRGs is supplemented with a survey of Mg\,II
absorbers in the vicinity of photometrically identified LRGs in the
SDSS data archive.  This survey allows us to utilize the vast survey data 
available in the SDSS archive to derive additional
constraints on the covering fraction of cool gas in $10^{13} \hmsol$
halos.  We adopt a $\Lambda$ cosmology with $\Omega_{\rm M}=0.3$ and
$\Omega_\Lambda = 0.7$, and a dimensionless Hubble parameter $h
=H_0/(100 \ {\rm km} \ {\rm s}^{-1}\ {\rm Mpc}^{-1})$ throughout the
paper. All distances are in physical units unless otherwise stated.

\section{Experiments}

To examine the cool gas content in massive halos, we have designed two
experiments.  The first one is a spectroscopic follow-up study of photometrically 
identified LRGs located near Mg\,II absorbers (Sample A below). The spectroscopic
observations allow us to identify physically associated LRG--Mg\,II absorber pairs
for interpreting the small-scale clustering signal seen in \citet{gauthier2009a}.  
In addition, the optical spectra offer additional knowledge for the stellar and ISM
properties of the LRGs.  The second experiment is a survey of Mg\,II
absorbers in the vicinity of photometrically identified LRGs (Sample B below). This
study allows us to make use of the vast amount of galaxy and QSO
survey data already available in the SDSS archive and derive
statistically significant constraints for the covering fraction of
cool gas around LRGs.

\subsection{Sample A -- LRGs in the vicinity of known Mg\,II absorbers}

The purpose of the spectroscopic study of LRGs is to establish a physical 
association between the cool gas traced by the Mg\,II absorbers and the dark matter halos 
of the LRGs. The LRGs are selected based on the photometric redshifts and typical 
photometric redshift uncertainties for these galaxies is  $\sigma_z = 0.03(1+z_{\rm phot})$ 
for $i'=19$ \citep{collister2007a}. The identification of physical pairs requires precise spectroscopic 
redshifts for the galaxies. 

To achieve this goal, we established a catalog of close Mg\,II-LRG pairs for follow-up 
spectroscopy by cross-correlating the \citet{prochter2006a} Mg\,II absorber catalog with the 
\citet{collister2007a} MegaZ-LRG sample.  The Mg\,II catalog is an extension of 
the SDSS DR3 sample to include DR5 QSO spectra. This sample of MgII absorbers 
has a 95\% completeness for absorbers of $W_r(2796) > 1$ \AA\ . The catalog 
contains 11,254 absorbers detected at $z_{\rm Mg\,II}=0.37-2.3$ along 9,774 
QSO sightlines. We excluded absorbers within 10,000 \kms\  from the QSO redshift
to avoid associated absorbers of the QSO. We excluded absorbers with 
$z_{\rm Mg\,II} <$ 0.4 or $z_{\rm Mg\,II}>0.7$ to match the redshift interval of the LRGs. 
A total of  2461 Mg\,II absorbers satisfied our selection criteria.  

The MegaZ-LRG catalog is a photometric redshift catalog of approximately $10^6$ 
LRGs found in the SDSS DR4 imaging footprint. The catalog covers more than 5,000 
deg$^2$ in the redshift range $0.4$ $<$ $z$ $<$ $0.7$. When cross-correlating 
the two catalogs, we rejected LRGs located within the photometric redshift uncertainty 
$3 \times \sigma_z$  from 
the QSO redshift. We selected only those LRGs that are in the foreground of the QSO.  

We found a total of 646 Mg\,II-LRG pairs within projected separation $\rho =$ 400 (physical) \hkpc. 
Note that the sample of 64 close Mg\,II-LRG pairs with $\rho \apll 300$ \hkpc found in 
\citet{gauthier2009a} is included in this larger sample of  646 pairs. From this sample, 
we only kept isolated LRGs.\footnote{ An isolated LRG is one that does 
not have a a neighboring LRG within the volume defined by their virial radius and photometric 
redshift uncertainty. ÊThis approach allow us to reduce the ambiguity of attributing the MgII 
absorber to one or more LRGs. ÊWe address possible contamination due to 
blue satellite galaxies in \S 5. The treatment of groups is
beyond the scope of this letter and will be discussed in a forthcoming
paper (Gauthier et al.\ 2010 in preparation)}.  Our final catalog for follow-up galaxy spectroscopy 
consisted of 331 'isolated' LRGs that are near a \emph{known} Mg\,II absorber. We call this sample 
\emph{Sample A}.  

The primary utility of Sample A was to understand the small-scale clustering 
signal seen in \citet{gauthier2009a}. The sample of spectroscopically confirmed physical 
LRG--Mg\,II pairs also allows us (1) to examine possible correlations between the stellar 
properties of the LRGs and the presence of cool gas at large radii, and (2) 
to study the kinematics of cool gas in massive halos with follow-up echelle spectroscopy of the 
absorbers. 

\subsection{Sample B -- random LRGs near QSO sightlines}

To derive constraints on the covering fraction of cool gas, we first
established a QSO--LRG pair sample by cross-correlating the
\citet{schneider2007a} QSO catalog and the MegaZ LRG catalog
\citep{collister2007a}).  The \citet{schneider2007a} catalog contains
more than 77K QSOs in the SDSS DR5 archive.  MegaZ is a photometric
redshift catalogue of approximately $10^6$ LRGs in the redshift range
0.4 $<$ $z_{\rm ph}$ $<$ 0.7.  The redshift uncertainties are
estimated to be $\sigma_{z}/(1+z_{\rm ph}) = 0.03$.

We considered QSOs that are in the distant background from the LRGs
with ($z_{\rm QSO}-z_{\rm ph}$) $>$ $3\,\sigma_{z}$ in order to
exclude correlated QSO--LRG pairs.  In addition, we included only QSOs
with $g'_{\rm QSO}<18.5$ and at $z \le 1.45$, in order to obtain
sensitive limits for $W_r(2796)$ and to remove contaminations due to
C\,IV absorption features.  Furthermore, we excluded LRGs found in
groups of two or more neighbors.  Finally, we considered
only pairs with $\rho < 400$ \hkpc\ at $z_{\rm ph}$, a maximum separation
that is about the expected size of the host dark matter halos.
This procedure yielded a total of 620 LRG--QSO pairs. We call this pair sample \emph{Sample B}. 
This catalog contains no \emph{a priori} information about the presence 
of Mg\,II absorber in the QSO spectra. It was designed to compute the covering fraction 
of cool gas in the dark matter halos of LRGs (see the analysis in \S\ 4.2). 

\section{Observations and Data Reduction}

In May 2009, we obtained medium-resolution spectra of three 
LRGs in Sample A (all within  $\rho = 160$ \hkpc of a known Mg\,II absorber) using the 
Double Imaging Spectrograph (DIS) on the 3.5~m telescope at the 
Apache Point Observatory. The blue and red cameras have a pixel 
scale of 0.4" and 0.42" per pixel, respectively, on the 3.5 m telescope. 
We used the B400/R300 grating configuration with a 1.5" slit. The B400 grating 
in the blue channel has a dispersion of 1.83 \AA\ per pixel and the R300 
grating in the red channel has a dispersion of 2.31 \AA\ per pixel. The blue 
and red channels with the medium resolution gratings together offer contiguous 
spectral coverage from $\lambda=3800$ \AA\ to $\lambda = 9800$ \AA\ and a 
spectral resolution of FWHM $\approx$ 500 \kms at $\lambda$= 4400 \AA\ and 
FWHM $\approx$ 400 \kms at $\lambda=7500$ \AA\ .The observations were 
carried out in a series of two exposures of between 1200 s and 1800 s each, and 
no dither was applied between individual exposures. Flat-field frames were taken 
at the end of the night for each night. Calibration frames for wavelength solutions 
were taken immediately after each science exposure using the truss lamps on the 
secondary cage. The DIS observations are characterized by a typical seeing of 1.3"
(as measured with the slitviewer guide star). Note that we also retrieved the spectrum 
of one LRG in Sample A (SDSSJ113731.00+060748.6)  from the SDSS 
archive and included it in our spectroscopic studies. 

We pursued our spectroscopic observations of LRGs 
in September 2009 using the Boller \& Chivens (B\&C) spectrograph on 
the du Pont 2.5-m telescope at Las Campanas Observatory, Chile. We 
obtained long-slit medium-resolution spectra for seven additional LRGs in Sample A.
In addition, we selected four LRGs from sample B for the spectroscopic B\&C observations. The 
B\&C camera has a plate scale of 0.7" per pixel. We used the 300 ln/mm grating with 
central wavelength of 6500 \AA\ and grating 
angle of 6 degrees. Since most spectral features of early-type galaxies at $z\sim 0.5$ 
are located at $\lambda > 5000$ \AA\ , we employed the blocking filter GG4495 to block 
the light blueward of $\approx$ 5000 \AA\ , in order to avoid second-order contamination 
from the blue part of the spectrum. The slit width for all observations 
was 1.1".  The adopted grating configuration had a dispersion of 3 \AA\ per pixel and 
a wavelength coverage of more than 6000\AA\ . 
For the calibration frames, we followed the methodology employed for the DIS 
observations. A series of dome and sky flats were taken each afternoon prior to the 
observations and calibration frames for wavelength solution were taken immediately 
after each science exposure. The typical seeing measured with the guide star was 
1-1.4" during most observations.

All of the DIS and B\&C spectroscopic data were reduced using standard long-slit 
spectral reduction procedures.  The spectra were calibrated to vacuum
wavelengths, corrected for the heliocentric motion, and
flux-calibrated using a spectrophotometric standard.  Redshifts of the
LRGs were determined based on a cross-correlation analysis using known
SDSS templates. The typical redshift uncertainty is $\Delta z \sim
0.0003$.

In summary, the spectroscopic sample consists  of 11 galaxies  in Sample A and four in Sample B. 
A journal summarizing the observations of the 15 galaxies can be found in Table 1. 

\section{Results}

\subsection{Optical spectra of LRGs}
The optical spectra described in \S\ 3 confirmed that five of the LRGs occur within 
350 \kms\ of a Mg\,II absorber. Four of these LRGs belong to sample A and one 
to sample B. Figure 1 shows the optical spectra and SDSS images of these five LRGs. 
The close proximity of these LRGs and Mg\,II absorber pairs 
($\rho \apll$ 300 \hkpc and $|\Delta v| \apll 350$ \kms) strongly argues for 
a physical association between the galaxy and the absorbing gas. We note that 
the volume spanned by $|\Delta v|=\pm 350$ \kms\ and $\rho = 350$ \hkpc\ is $\sim$ 
5 co-moving (\hmpc)$^3$ at $z=0.5$. Within this small volume, we estimate based on the 
best-fit luminosity functions determined for red galaxies at $z \sim 0.5$ ($M_*- 5\log~h=-19.8$;
$\phi_*=5.16\times 10^{-3}$ (\hmpc)$^{-3}$; see e.g.,\ \citealt{brown2007a,brown2008a}) that the probability of the LRG being a random galaxy is negligible. 


\begin{centering}
\begin{deluxetable*}{llrrrcccc}
\tablecaption{Summary of the Long-slit Spectroscopic Observations of Luminous 
Red Galaxies}
\tablewidth{0pt}
\tablehead{\colhead{ID} & \colhead{RA(J2000)} & \colhead{Dec(J2000)} & \colhead{$z_{\rm phot}$}
& \colhead{$i'$} & \colhead{Instrument} & \colhead{Exptime (sec)} & \colhead{UT date} & \colhead{Sample} }
\startdata
SDSSJ003816.29$-$092550.5 & 00:38:16.29 & -09:25:50.5 & 0.51 & 18.70 & B\&C & 2$\times$1800 & 2009-09-21 & B \\
SDSSJ011942.14$-$090225.4 & 01:19:42.14 & -09:02:25.4 & 0.49 & 19.88 & B\&C & 3$\times$2400 & 2009-09-18 & A\\
SDSSJ015452.46$-$095533.6 & 01:54:52.46 & -09:55:33.6 & 0.55 & 19.98 & B\&C & 2$\times$2400 & 2009-09-21 & B\\
SDSSJ021819.25$-$083331.8 & 02:18:19.25 & -08:33:31.8 & 0.56 & 19.46 & B\&C & 2$\times$2400 & 2009-09-19 & A \\
SDSSJ023705.35$-$075513.7 & 02:37:05.35 & -07:55:13.7 & 0.55 & 19.44 & B\&C & 1800 $+$ 900 &  2009-09-21 & B \\
SDSSJ034802.50$-$070339.3 & 03:48:02.50 & -07:03:39.3 & 0.49 & 19.08 & B\&C & 2$\times$2400 & 2009-09-18 & A \\
SDSSJ113731.00$+$060748.6 & 11:37:31.00 & +06:07:48.6 & 0.61 & 18.00 & SDSS & ... & ... & A \\
SDSSJ142610.27$+$594704.6 & 14:26:10.27 & +59:47:04.6 & 0.48 & 18.69 & DIS &  1800 $+$ 1500 & 2009-05-26 & A \\
SDSSJ160725.87$+$471221.7 & 16:07:25.87 & +47:12:21.7 & 0.56 & 19.70 & DIS & 2$\times$1800 & 2009-05-26 & A \\
SDSSJ161713.60$+$243254.8 & 16:17:13.60 & +24:32:54.8 & 0.57 & 19.06 & DIS & 2$\times$1800 & 2009-05-26 & A \\
SDSSJ204820.88$-$001640.1 & 20:48:20.88 & -00:16:40.1 & 0.45 & 19.47 & B\&C & 2$\times$2400 $+$ 1800 & 2009-09-19 & A \\
SDSSJ205145.36$-$052121.2 & 20:51:45.36 & -05:21:21.2 & 0.50 & 19.44 & B\&C & 2$\times$2400 & 2009-09-18 & A\\
SDSSJ211625.92$-$062415.4 & 21:16:25.92 & -06:24:15.4 & 0.55 & 19.31 & B\&C & 2$\times$2400 $+$ 2000 & 2009-09-21 & B \\
SDSSJ212713.92$-$000747.2 & 21:27:13.92 & -00:07:47.2 & 0.47 & 19.19 & B\&C & 2$\times$1800 & 2009-09-20 & A\\
SDSSJ232450.16$-$095048.4 & 23:24:50.16 & -09:50:48.4 & 0.53 & 19.79 & B\&C & 2$\times$2400 & 2009-09-19 & A
\enddata
\end{deluxetable*}
\end{centering}

\begin{figure*}
 \vspace{0.5pt}  \centerline{\hbox{ \hspace{0in}
      \includegraphics[angle=0,scale=0.40]{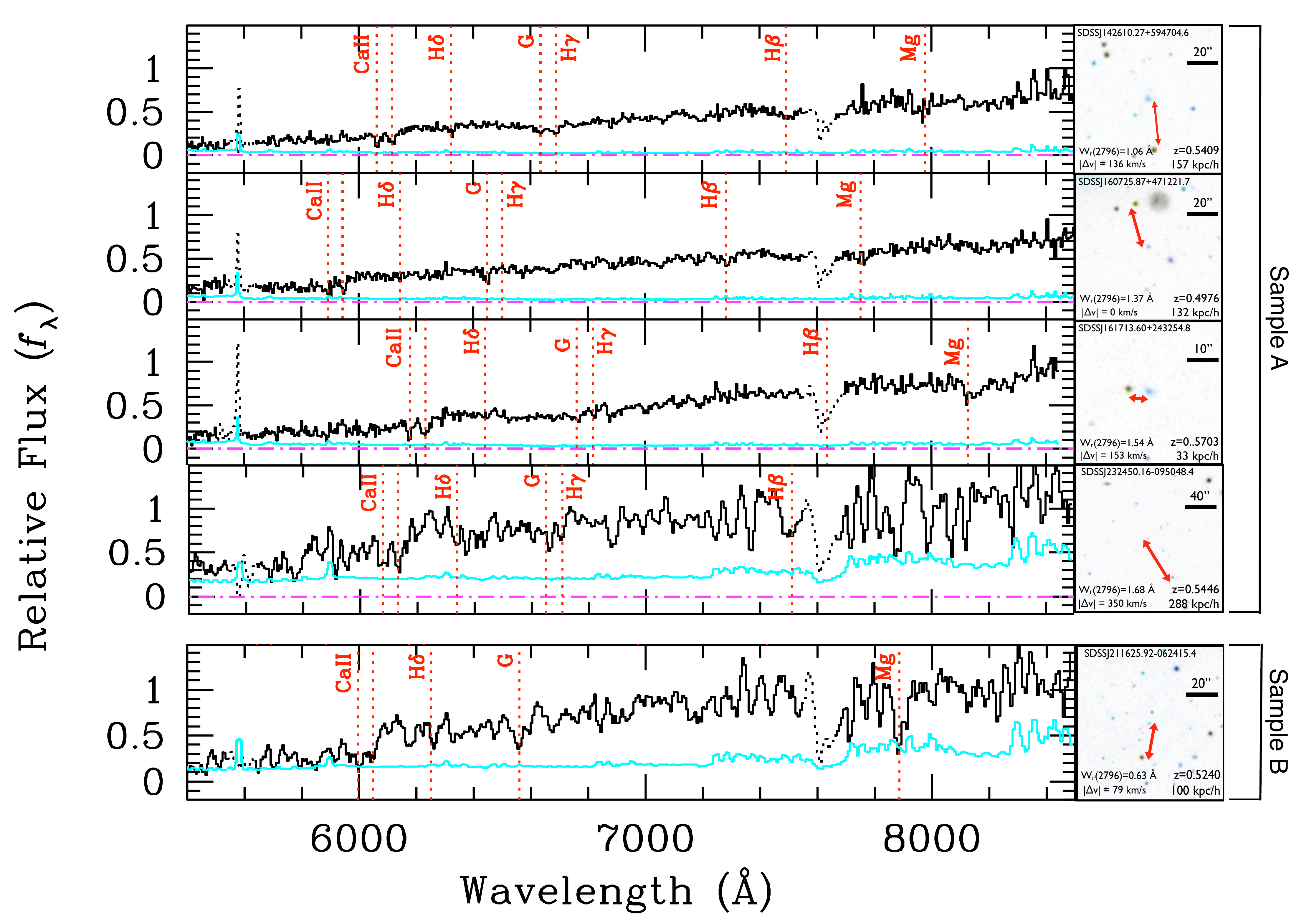}
      }
   }
\caption{Spectra of the LRGs in our spectroscopic sample with associated 
Mg\,II absorbers, along with the 
thumbnail images of the galaxies. The top four LRGs were selected from 
Sample A (\S\ 2.1) and the bottom one was selected from 
Sample B (\S\ 2.2). For each LRG, we show the reduced spectrum 
in thick solid histograms  and the corresponding
1-$\sigma$ error array in thin solid histograms.  The dotted features
are contaminating sky lines or artifacts.  The thumbnail images are reproduced from
the SDSS data archive to show the relative alignment of the LRG--QSO
pairs. The LRG is at the center of each image and is connected to the
background QSO by the arrow.  The redshift of the LRG and the
projected distance to the QSO sightline are listed in the bottom right
corner.  The velocity separation between the LRG and Mg\,II absorber 
is listed in the bottom left along with $W_r(2796)$.}
\label{spectra}
\end{figure*}

\begin{figure*}
 \vspace{0.5pt}  \centerline{\hbox{ \hspace{0in}
      \includegraphics[angle=0,scale=0.40]{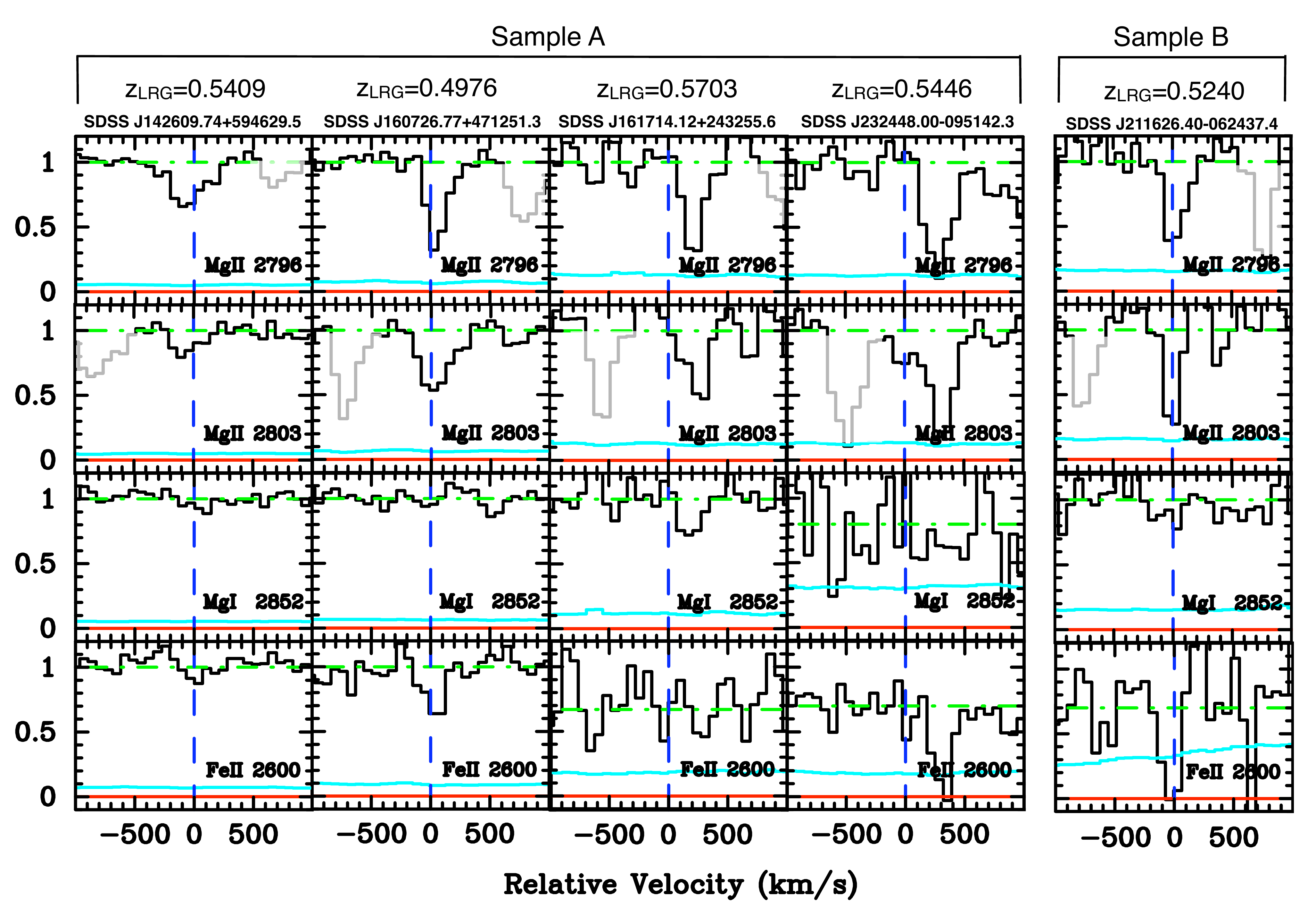}
      }
   }
\caption{Absorption profiles of the five physical LRG--Mg\,II absorber pairs 
in our spectroscopic sample of LRGs. The first four objects belong 
to Sample A and the last one to Sample B. We  included the associated 
Mg\,I $\lambda\,2852$ and Fe\,II $\lambda\,2600$ transitions. Zero velocity 
corresponds to the spectroscopic redshift of the associated LRG.  
Contaminating features are shown in gray.}
\label{vel}
\end{figure*}

In Figure \ref{vel} 
we present absorption profiles of the Mg\,II, Mg\,I and Fe\,II transitions from
the SDSS QSO spectra for these five physical pairs.
In each panel, zero velocity corresponds to the spectroscopic redshift
of the LRG.  All absorbers occur within
$| \Delta v | \apll 350$ \kms\ of the spectroscopically identified
LRGs.  Although the spectral resolution of SDSS spectra does not allow
us to resolve individual components, the panels show that three of the
Mg\,II absorbers have associated Fe\,II and Mg\,I absorption features.

All 15 LRG spectra exhibit spectral features dominated by absorption transitions, indicating an 
old underlying stellar population and little star formation in the recent past. A  stellar population 
synthesis analysis done with the \citet{bruzual2003a} spectral library established 
that the most likely stellar population ages for these LRGs span 1-11.75 Gyr 
(Gauthier et al.\  2010 in preparation).  

\subsection{Incidence of cool gas in LRG halos}

\begin{figure}
 \vspace{0.5pt}  \centerline{\hbox{ \hspace{0.0in}
	\includegraphics[angle=0,scale=0.40]{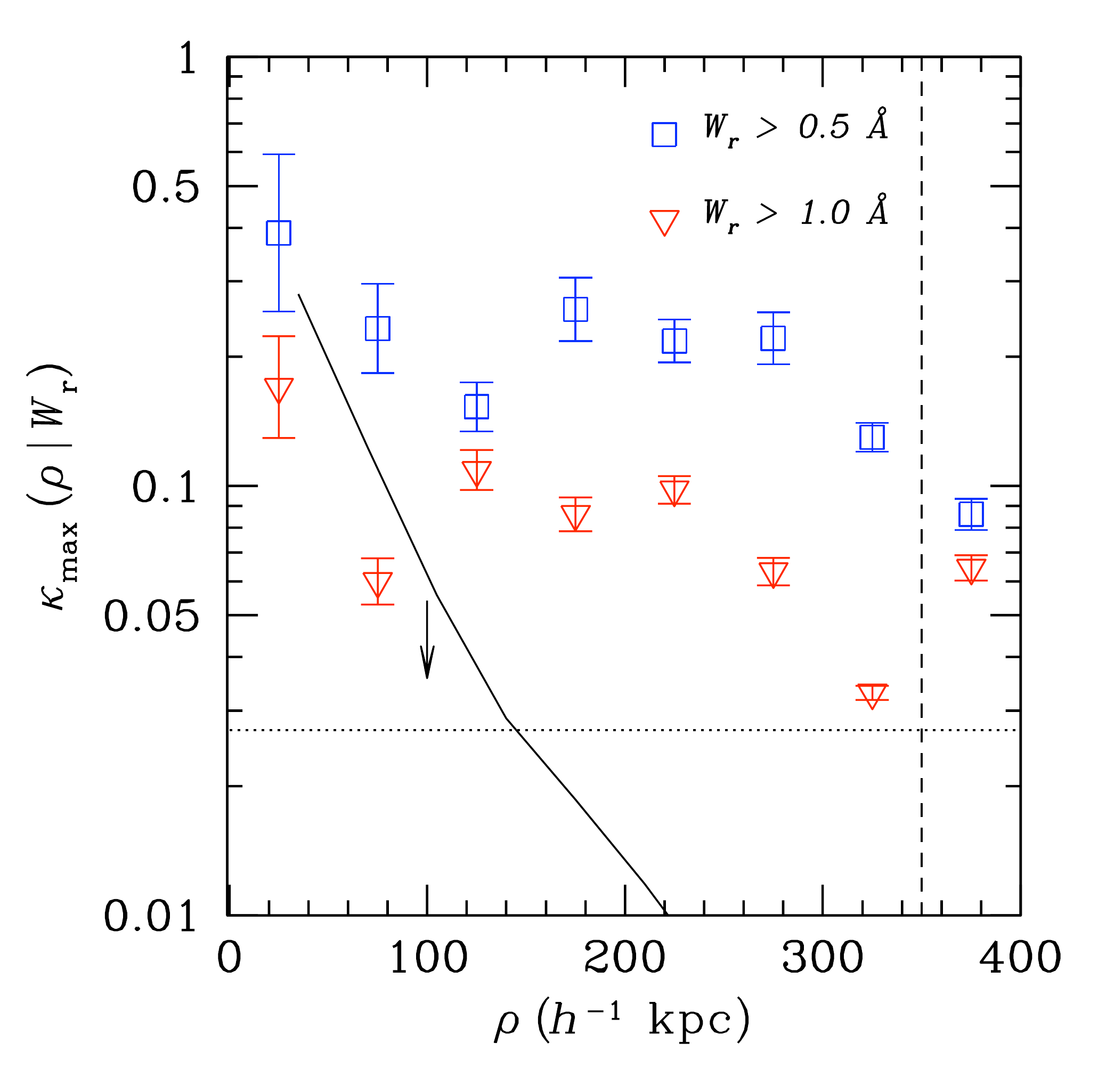}
      }
   }
\caption{The incidence of cool gas as probed by the presence/absence of
Mg\,II absorption features, $\kappa_{\rm max}$, at different projected 
radius of massive LRG halos in Sample B. The vertical dashed line represents the 
typical virial radius of the LRG population (see \citealt{gauthier2009a}).
Despite the large redshift uncertainties of the LRGs, the observed incidence 
of Mg\,II absorbers places a strong constraint on the maximum covering 
fraction of cool gas around LRGs.  We estimate the contribution due to 
random background absorbers that occur within the redshift interval of 
the LRGs by chance coincidence.  The result is shown in the dotted line. 
Including correlated absorbers due to large-scale matter clustering does 
not increase the background contribution to beyond 3\%.  The solid curve 
corresponds to the expected {\it maximum} contribution from satellite 
galaxies to the observed incidence of $W_r(2796)\ge 1.0$ \AA\ absorbers 
(see Discussion section).  Our study shows that neither satellite galaxies 
nor background random/correlated galaxies are sufficient to account for the 
observed incidence of Mg\,II absorbers in the vicinity of LRGs. Note that 
the errobars on $\kappa_{\rm max}$ are derived from the Poisson distribution.}
\label{up}
\end{figure}

We determined the covering fraction of cool gas in massive halos,
based on the presence/absence of Mg\,II absorbers in the spectrum of the 
background QSO near the redshifts of the LRGs  in Sample B.  
For each LRG--QSO pair, we visually inspected the QSO spectrum in search of Mg\,II absorbers in
the redshift uncertainty range of the LRG, $z_{\rm ph}\pm
3\,\sigma_z$.  For $\sigma_z \sim 0.045$ at $z_{\rm ph}=0.5$, this
interval corresponds to approximately $\pm 400$\AA\ centered at $\sim
4200$\AA\ . We defined this interval as the search interval, which is
large enough to identify all possible Mg\,II absorbers that may be
associated with the LRG.  Despite the large redshift uncertainties,
however, our search yielded a conservative maximum gas
covering fraction $\kappa_{\rm max}$ of Mg\,II absorbing gas in LRG
halos.  If no absorber was visually detected, we measured a 2-$\sigma$
upper limit per spectral resolution element, $W_{\rm lim}$, for the
equivalent width of the Mg\,II $\lambda\,2796$ transition, using the
noise spectrum.  For each sightline with a detection, we measured the
redshift and strength of the absorber based on a Gaussian profile
analysis.  For these sightlines, we also determined $W_{\rm {lim}}$
following the procedure described earlier.  Because the $S/N$ varies
substantially between QSO spectra, $W_{\rm lim}$ represents the
minimum absorption strength that can be recovered in each QSO spectrum. 
It serves to define a homogeneous sample of QSO spectra that can 
be adopted to determine the incidence of Mg\,II absorbers.  

To determine the incidence of Mg\,II absorbers with $W_r(2796)\ge W_0$
\AA\ around LRGs, we first identified from the parent LRG--QSO pair
sample the QSO spectra that satisfy $W_{\rm lim}<W_0$ \AA\ to ensure 
that the QSO spectra have sufficient $S/N$ for revealing a Mg\,II absorber 
of minimum strength $W_0$.  Next, we
determined $\kappa_{\rm max}$ by evaluating the fraction of the QSO
spectra that display a Mg\,II absorber of $W_r(2796)>W_0$ in the
vicinity of the LRGs.  We calculate $\kappa_{\rm max}$ for different
values of $W_0$ in different $\rho$ intervals.

This search allowed us to estimate the maximum possible
covering fraction of cool gas based on the incidence of Mg\,II
in LRG halos (see \S\ 5 for a discussion on the contamination 
rate due to correlated and random galaxies).  We first considered the
incidence of Mg\,II absorbers at different absorption threshold $W_0$.
All QSO spectra in the 620 close LRG--QSO pairs have sufficient $S/N$
($W_{\rm lim}<1$ \AA) for uncovering an absorber of $W_r(2796)\ge 1$
\AA.  Our search showed that 45 of the 620 LRGs have a $W_r(2796)\ge
1$ \AA\ absorber at $\rho < 400\ h^{-1}$ kpc, indicating a mean
covering fraction averaged over the entire LRG halos of $\kappa_{\rm
max}(\rho<400)=0.07$ for $W_r(2796)>1$ \AA\ absorbers.  Next, we 
calculated $\kappa_{\rm max}$ in different $\rho$ intervals.  Figure
\ref{up} shows that for $W_r \apg 1$\AA\ 
$\kappa_{\rm max}(\rho)\approx 10$\% at $\rho < 300\ h^{-1}$ kpc but 
declines to $\kappa_{\rm max}\approx 5$\% at larger $\rho$ (triangles).

Similarly, 575 QSO spectra in the 620 close LRG--QSO pairs have
sufficient $S/N$ ($W_{\rm lim}<0.5$ \AA) for uncovering an absorber of
$W_r(2796)\ge 0.5$ \AA, and 105 of the 575 LRG members have a
$W_r(2796)\ge 0.5$ \AA\ absorber at $\rho < 400\ h^{-1}$ kpc.  The
search therefore yielded a mean covering fraction averaged over the
entire LRG halos of $\kappa_{\rm max}(\rho<400)=0.18$ for absorbers of
$W_r(2786)>0.5$ \AA.  The estimated $\kappa_{\rm max}$ versus $\rho$
for $W_r(2796)\ge 0.5$ \AA\ absorbers are also presented in Figure
\ref{up} (squares).

The $\kappa_{\rm max}$ results at small separations ($\rho \apll 100$ \hkpc) 
are consistent with the covering fraction measurements from the spectroscopic 
Sample B. We found 1/4 LRGs physically associated with the Mg\,II absorber and all four 
pairs are within $\rho = 100$ \hkpc. We are currently increasing the number of 
spectroscopic pairs from Sample B to obtain a more accurate measurement of $\kappa$. 

\section{Discussion}

Our follow-up spectroscopic study of close LRG--Mg\,II pairs has
confirmed a physical association between the Mg\,II absorber and the
LRG in five of the 15 cases studied.  The small velocity
separations, $|\Delta\,v|\apll 350$ \kms\ and projected distances $\rho
< 160\,h^{-1}$ kpc demonstrate that at least some of the strong Mg\,II
absorbers [$W_r(2796)>1$ \AA ] that contributed to the small-scale
[$r_p=(1+z)\,\times \, \rho<600\,h^{-1}$ kpc] clustering signal in Gauthier et
al. (2009) do indeed originate in the hosting dark matter halos of
LRGs.  According to a stellar population analysis (using the stellar 
templates from \citealt{bruzual2003a}) done on the LRG spectra, we found that 
these galaxies are characterized by a stellar population at least 1Gyr old 
with a most likely age between 1-11.75 Gyr. The optical spectra are 
characteristic of an evolved stellar population with little recent star formation. 

In addition, our search of Mg\,II absorption features in the vicinity
of LRGs has yielded a conservative estimate of the maximum covering
fraction of cool gas in LRG halos.  Because of the redshift
uncertainties of the LRGs, we cannot distinguish between whether the
observed Mg\,II absorbers are physically associated with the LRG halos
or these absorbers occur due to chance coincidence, correlated
galaxies, or surrounding satellite galaxies.  Despite these various
uncertainties, the observed incidence of Mg\,II absorbers places a
strong constraint on the maximum covering fraction of cool gas in
massive LRG halos as probed by the Mg\,II absorbers.  We find that the
cool gas covering fraction of absorbers with $W_r(2796)\ge 0.5$ \AA\
in massive LRG halos is no more than 20\% and no more than 7\% for
stronger absorbers $W_r(2796)\ge 1$ \AA.

Although to establish a physical association between these close
LRG--Mg\,II absorber pairs requires spectroscopic data of the LRGs, we
assess possible contaminations due to surrounding satellite galaxies
and background random/correlated absorbers based on theoretical
expectations of the satellite population and the mean number density
of Mg\,II absorbers per line of sight observed along random QSO
sightlines.

To estimate the incidence of Mg\,II absorbers due to satellite
galaxies in the LRG halos, we first adopt the mass function of
subhalos (Equation 10 in \citealt{tinker2009c}).  The mass of subhalos
is defined as the mass at the time of accretion---this mass will
correlate with the gaseous halo that the subhalo initially had.  Next,
we adopt the gas radius at $W_r(2796) \geq 1$ \AA\ and the covering
fraction within the gas radius as a function of halo mass from
\citet{tinker2010a}.  The total cross section of subhalos within host halos of
mass $10^{13}$ \hmsol\ is then estimated according to

\begin{eqnarray}
{\hat \kappa}_{\rm sub}(M_{\rm host}) &=& \frac{1}{\sigma(M_{\rm host})} \int dM_{\rm sub} n(M_{\rm sub}|M_{\rm host}) \\ \nonumber
{} & &\times {} \pi R^2(M_{\rm sub}, W_r=1) \kappa_g(M_{\rm sub})
\label{kappa_eqn}
\end{eqnarray}
where $M_{\rm host}$ is the host halo mass (in this case, $10^{13}$
\hmsol), $M_{\rm sub}$ is subhalo mass at time of accretion, $n(M_{\rm
sub}|M_{\rm host})$ is the subhalo mass function from
\citet{tinker2009c}, $R$ is the gas radius for $W_r(2796)=1$ \AA\, and
$\kappa_g$ is the covering fraction within the gas radius.  Both $R$
and $\kappa_g$ are adopted from \citet{tinker2010a}, which are
calibrated on data to match the frequency function of absorbers and
large scale bias as a function of $W_r(2796)$.

Under the assumption that the gaseous halos of satellite halos are the
same as field halos of similar mass, Equation 1 yields a maximal
covering fraction of 1.8\% at $\rho \apll 350$ \hkpc.  This is the
maximum value because tidal stripping and ram pressure are not taken
into account (e.g.,\ \citealt{gunn1972a,balogh2000a}).  Equation 1 is also
maximal because it assumes that the subhalos have a random
distribution within the parent halo. If we further assume that the
subhalos follow an NFW profile (Wetzel \& White 2009), ${\hat
\kappa}_{\rm sub}$ depends on impact parameter as shown in Figure
4. The satellite contribution is at most comparable to the
$\kappa_{\rm max}$ measurements at $\rho < 100$ \hkpc, but the
satellite covering factor of cool gas is $\apg 1$ dex below
$\kappa_{\rm max}$ at $\rho > 200$ \hkpc. 

To estimate the incidence of Mg\,II absorbers due to structure along a line 
of sight that intersects a large halo, we adopt the observed mean number density
of $W_r(2796)\ge 1$ \AA\ absorbers from Nestor \etal\ (2005) and
Prochter \etal\ (2006).  These authors report a number density of
$n\approx 0.1$ per unit redshift interval per line of sight at $z=0.5$
for Mg\,II absorbers of $W_r(2796)\ge 1$ \AA.  Over a redshift
interval of $\Delta\,z=0.27$, which corresponds to the redshift
uncertainty of the LRGs, we therefore expect a background
contamination of 0.027 absorbers per line of sight.  To estimate the
incidence of Mg\,II absorbers due to correlated large-scale galaxy
distribution, we calculate the predicted incidence of correlated $L_*$
halos around the LRGs. We select $L_*$ halos because they 
are expected to be the dominant contaminant (\citealt{tinker2008a,tinker2010a}).
First, we adopt Equation (11) in \citet{tinker2010a} and 
modify this equation to take into account the correlated structures along the line 
of sight. We integrate along the line-of-sight distance from $-300$ to 
$+300$  \hmpc. We repeat the calculation for each projected distance to the LRG
and determine the number of halos that could contribute to the signal based on their mass and size. 
 Our calculation indicates roughly 10\%
enhancement in the background number density of Mg\,II absorbers per
line of sight.  We therefore conclude that background random and
correlated galaxies would have contributed no more than 3\% of the
observed Mg\,II absorbers around the massive LRGs.

Our exercise shows that contaminations due to surrounding satellite
galaxies and background random/correlated galaxies are insufficient to
account for the observed incidence of Mg\,II absorbers in the vicinity
of LRGs.  The maximum covering fraction of strong absorbers is $\sim$ 7\% averaged 
over the entire halo. 

\citet{bowen2006a} reported 100\% covering fraction of Mg\,II absorbers within 
100 kpc projected distance from four foreground QSO at $z=0.65-1.55$. The observed
large incidence of Mg\,II absorbers in QSO halos appears to be inconsistent with the 
low gas covering fraction found in the LRG halos. This discrepancy can be understood 
by considering the mean mass scales of QSOs and LRGs. \citet{ross2009a} calculated the 
clustering of SDSS spectroscopic QSOs at $z<2.2$. They found that, at $z\sim 0.5$, the bias 
of the dark matter halos traced by QSOs is $b\sim1.4$ which is smaller than the bias estimate
for the LRGs at the same redshift ($b\sim2$; \citealt{gauthier2009a}). This implies that 
the typical mass of the dark matter halos hosting QSOs is around $\log$ M/$h$ $\sim$ 12.3 (see 
figure 12 in \citealt{ross2009a}), which is about one dex smaller than what is found for 
LRGs in \citet{gauthier2009a}. The difference in the observed gas covering fraction therefore 
reflects the differences in the cold gas content of halos on different mass scales (see e.g.,\
\citealt{tinker2008a}, Chen et al. 2010 - ApJ submitted).

Recall that Mg\,II absorbers trace high column density gas of neutral
hydrogen column density $N({\rm H\,I})> 10^{18}$ \cmjj\
\citep{bergeron1986a, rao2006a}.  Furthermore, Rao et al.\ (2006) have
shown that on average 42\% of $W_r(2796)>0.6 $ \AA\ Mg\,II absorbers
with associated strong Fe\,II and Mg\,I absorption transitions are
likely to contain neutral gas of $N({\rm H\,I})> 2\times 10^{20}$
\cmjj.  Two of the Mg\,II absorbers presented in Figure 2
satisfy these criteria and are therefore likely damped \lya\
absorbers.  Our study therefore provide direct empirical evidence for
the presence of high column density cool gas in $10^{13}$ \hmsol\
halos.  A large spectroscopic sample of LRGs will not only improve the
contraints for $\kappa_g$ in massive halos but also provide additional
insights into the origin of the cool gas.

\acknowledgments We thank the anonymous referee for helpful comments 
that improved the draft. We also thank Jennifer Helsby for assistance in obtaining
and reducing some of the LRG spectra and Michael Rauch for helpful 
comments during the preparation of the draft.  JRG acknowledges support from
the Sigma-Xi Grant-in-Aid of Research program and from the Brinson
Foundation predoctoral fellowship.  HWC acknowledges partial support
from NASA Long Term Space Astrophysics grant NNG06GC36G and an NSF
grant AST-0607510.

\bibliographystyle{apj}
\bibliography{ms10}

\clearpage

\end{document}